\documentclass[sigconf, screen, review]{acmart}

\usepackage{style}

\AtBeginDocument{%
  }

\setcopyright{none} 
\copyrightyear{2026}
\acmYear{2026}
\acmDOI{XXXXXXX.XXXXXXX}

\acmConference[MICRO 2026]{The 58th IEEE/ACM International Symposium on Microarchitecture}{October 31--November 04, 2026}{Athens, Greece}

\acmISBN{978-X-XXXX-XXXX-X/XX/XX}

\begin{document}

\title{Fast and Fusiest: An Optimal Fusion-Aware Mapper for Accelerator Design}

\author{Tanner Andrulis}
\authornote{These authors contributed equally to this work.}
\email{andrulis@mit.edu}
\affiliation{%
  \institution{MIT}
  \city{Cambridge}
  \state{MA}
  \country{USA}
}

\author{Michael Gilbert}
\authornotemark[1]
\email{gilbertm@mit.edu}
\affiliation{%
  \institution{MIT}
  \city{Cambridge}
  \state{MA}
  \country{USA}
}

\author{Vivienne Sze}
\email{sze@mit.edu}
\affiliation{%
  \institution{MIT}
  \city{Cambridge}
  \state{MA}
  \country{USA}
}

\author{Joel S. Emer}
\email{emer@csail.mit.edu}
\affiliation{%
  \institution{MIT / Nvidia}
  \city{Cambridge}
  \state{MA}
  \country{USA}
}





\begin{abstract}

A low-latency and energy-efficient tensor algebra accelerator design must optimize how data movement and operations are scheduled (i.e., mapped) in the accelerator architecture. A key mapping optimization is fusion, meaning holding data on-chip between computation steps in the workload, which has been shown to reduce energy and latency by reducing expensive off-chip data movement. However, the optimal fusion choice depends on the workload and workload shape, and a mapper, which searches for the optimal mapping, can improve energy and latency significantly. However, prior mappers cannot find optimal mappings with fusion (i.e., fused mappings) in a feasible runtime because the number of fused mappings to search increases exponentially with the number of computation steps in the workload.

In this paper, we introduce the Fast and Fusiest Mapper (FFM), a mapper to quickly find optimal mappings in a comprehensive fused mapspace for tensor algebra workloads. FFM shrinks the search space by pruning subsets of mappings (i.e., partial mappings) that are shown to never be a part of optimal mappings, quickly eliminating all suboptimal mappings containing those partial mappings. Then FFM joins partial mappings to construct optimal fused mappings. Using FFM, we demonstrate an energy-delay-product (EDP) reduction by up to $1.8\times$ compared to TransFusion, a state-of-the-art accelerator with hand-optimized fusion. Moreover, we show that FFM finds mappings orders of magnitude faster ($>10,000\times$) than prior automated mappers TileFlow and SET, and given the same runtime, reduces EDP by $>2\times$.
\end{abstract}

\maketitle
\section{Introduction}\label{sec:introduction}

The latency and energy of a tensor algebra accelerator depend on how data movement and operations are scheduled in the accelerator architecture (\ie the \emph{mapping}~\cite{timeloop}). Various mapping optimizations have been proposed to leverage various features of the workload to lower latency and energy, including dataflow~\cite{eyeriss_isca,flat,fusemax}, tiling~\cite{ruby,looptree,tileflow}, and fusion~\cite{looptree,flat,fusemax,transfusion,optimus}. Often, state-of-the-art accelerators combine several of these optimizations~\cite{fusemax, transfusion, flat}.

This paper, like many prior works, uses the \emph{Einsum} notation~\cite{teaal, edge, extensor, fusemax} to analyze and reveal the space of optimizations for computation steps (which we will also call \emph{Einsums}) in a workload.

Mapping optimizations often are aimed at reducing high-energy, high-latency data movement both on and off-chip. To accomplish this, we can \emph{reuse} data across multiple computations rather than fetching it from more costly memories. Within each Einsum, we can capture \emph{intra-Einsum reuse} by optimizing computation order (\ie \emph{dataflow}) and data tiling~\cite{timeloop,eyeriss_isca,maestro}. Between Einsums, we can capture \emph{inter-Einsum} reuse by \emph{fusing} to share data across multiple Einsums (\eg reusing the intermediate result between a matrix multiplication and a following softmax). Many workloads, such as Transformer self-attention, contain Einsums with especially low intra-Einsum reuse opportunity, and fusion has become a critical optimization for those workloads~\cite{flat,fusemax,flashattention,flashattention2}.

However, a challenge that state-of-the-art accelerators have not adequately addressed is that mapping optimizations often involve tradeoffs between these reuse types, and the optimal choice changes depending on the workload and architecture. Even minor workload changes (\eg changing the number of input tokens of a Transformer) can result in drastically different optimal mappings. For example, our analysis in Section~\ref{sec:result:comprehensive_mapspace} shows that for long sequence lengths, TransFusion~\cite{transfusion}, a state-of-the-art accelerator for Transformers, often incurs high energy by fusing too many Einsums. The reason is that leveraging intra- and inter-Einsum data reuse both require reserving on-chip buffer capacity. With shorter sequence lengths, there is less intra-Einsum reuse opportunity (\eg each weight is reused for fewer tokens), and intermediate data is smaller since it scales with the number of tokens. These factors mean allocating more on-chip buffer capacity to make fusion beneficial. However, the opposite is true with longer sequence lengths.





A solution to this challenge is to use a \emph{mapper}~\cite{timeloop}, which searches for an optimal mapping from a space of mapping choices (\ie a \emph{mapspace}~\cite{timeloop}) for a given architecture and workload, but \blfootnote{\label{url} FFM is available at \emph{<to be posted after peer review.>}}
exploring mappings with fusion is particularly challenging because it requires co-optimizing the mappings of multiple Einsums. For a mapper to find more efficient mappings than current solutions, a \emph{comprehensive} mapspace must be considered (\ie various dataflow, tiling, and fusion choices). However, in such a comprehensive mapspace, there are many dataflow and tiling choices for each Einsum, and the number of overall mappings is the product of the number of choices for each Einsum. This makes the number of mappings to consider increase exponentially with the number of Einsums in a workload.

\newcommand{\ml}[1]{\begin{tabular}{c}#1\end{tabular}}
\renewcommand\theadfont{}
\begin{table}
    \centering
    \setlength{\tabcolsep}{4pt}
    \begin{tabular}{lccc}
         \thead{\textbf{Mapper}} & \thead{\textbf{\makecell{Comprehensive\\mapspace}}}  & \thead{\textbf{Optimal}} & \thead{\textbf{\makecell{Time to \\ optimal}}} \\
    \toprule
        \ml{Timeloop\cite{timeloop}\\Maestro\cite{maestro}\\Zigzag\cite{zigzag}} & \bad[\ml{\textbf{No}:\\No Fusion}] & \bad[No] & N/A \\
        \hline
        \ml{Optimus\cite{optimus}\\ConvFusion\cite{convfusion}\\FusedCNN\cite{fusedcnn}\\DeFiNES~\cite{defines}\\FLAT~\cite{flat}} &
        \bad[\ml{\textbf{No:}\\Limited dataflows}] &
        \bad[No]
        & N/A\\
        \hline
        \ml{SET~\cite{set}\\Tileflow~\cite{tileflow}} & \good[Yes] & \limited[Near*] & \bad[Slow*] \\
        \hline
        \textbf{This work}    & \good[Yes] & \good[Yes] & \good[Fast**] \\
    \bottomrule

    \multicolumn{4}{l}{\footnotesize{*Within $1-10\%$ of optimal after approx. 15,000 CPU hours (Section~\ref{sec:results}).}} \\
    \multicolumn{4}{l}{\footnotesize{**Optimal after approx. 1.5 CPU hours (Section~\ref{sec:results}).}} \\
    \end{tabular}
    \caption{Only this work is fast and optimal.
    }
    \label{tab:mapper_comparison}
\end{table}

Current mappers~\cite{tileflow, set} inadequately address this large mapspace size using black-box search algorithms (\eg genetic algorithms~\cite{tileflow} or simulated annealing~\cite{set}), which can be slow or even fail to converge to an optimal mapping (see row 3 of Table.~\ref{tab:mapper_comparison}). In Section~\ref{sec:results}, we show that these works require $\geq15,000$ CPU hours ($\approx2$ CPU years) to converge to within 1\% of optimal energy-delay-product for a reasonably-sized workload (a transformer layer with 10 Einsums). Meanwhile, other mappers do not support a comprehensive mapspace, often supporting only a particular dataflow, tiling, or fusion (see rows 1 and 2 of Table~\ref{tab:mapper_comparison})~\cite{timeloop, maestro, zigzag, defines, convfusion, fusedcnn, optimus,flat,flashattention}.

In this paper, we introduce the \ffmlong (\ffm), a fast mapper that can find optimal\footnote{While the optimal definition is mapspace-dependent and we cannot predict future mapping optimizations, our mapspace is a superset of the mapspaces in many prior, and state-of-the-art works ~\cite{timeloop,maestro,zigzag,optimus,convfusion,fusedcnn,defines,flat,set,tileflow}.}
fused mappings in a comprehensive mapspace for a user-configured architecture and workload. \ffm builds a mapping by starting from partial mappings (\ie \emph{pmappings}) that each map only one Einsum. \ffm then iteratively joins pmappings until a mapping for the full workload is created. At each step, we group and prune pmappings to keep the search space small and let us quickly find an optimal mapping.

\begin{figure}
    \centering
    \includegraphics[width=0.9\linewidth]{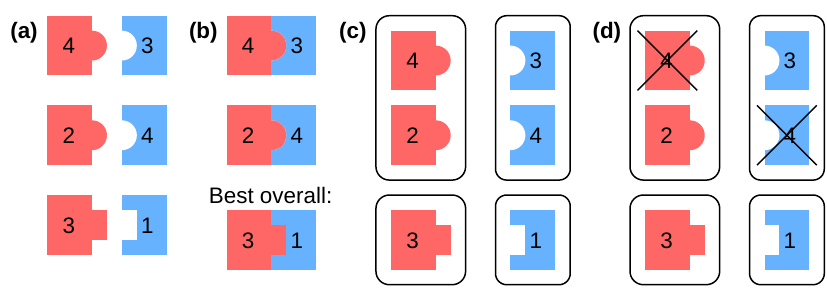}
    \caption{An analogy of pmappings as puzzle pieces. (a) Tabs and sockets represent compatibility. Numbers represent a cost metric. (b) Pieces can be joined in several ways. (c) Pieces can be grouped by tabs/sockets and (d) pruned per-group.}
    \label{fig:puzzle}
\end{figure}

To provide intuition about this process, Fig.~\ref{fig:puzzle} shows a simplified analogy with a sequence of different colored puzzle pieces (Einsums) to join together and two groupings for each color piece based on their tabs and sockets. We are constructing a jigsaw puzzle (a full mapping) from a collection of puzzle pieces (pmappings). The full puzzle (full mapping) must have exactly one piece of each color (one pmapping for each Einsum). Pieces (pmappings) can only be joined if tabs and sockets match (pmappings are \emph{compatible} if the ways they exchange shared data match). Our objective is to minimize the sum of the numbers on the pieces (some measure of cost, \eg energy of a pmapping). Fig.~\ref{fig:puzzle} (b) shows that the best combination of puzzle pieces is ``red 3" and ``blue 1." We can find this solution by checking all $3\times3=9$ combinations. But as we increase the number of colors of pieces in the puzzle (number of Einsums in the workload), the runtime of this brute-force approach would grow exponentially, and this approach becomes infeasible.

Now, we apply two optimizations to reduce the effort to find the optimal solution. \textbf{First, we can avoid trying many incompatible pairs by grouping pieces with the same tab and socket shapes and only considering combinations within compatible groups.} As Fig.~\ref{fig:puzzle}(c) shows, this optimization reduces the possible combinations from 9 to $2\times2+1\times1=5$. \textbf{Second, note that we only have to consider the best piece, with the lowest number, in each group.} Fig.~\ref{fig:puzzle}(d) shows that we can maintain overall optimality while greedily selecting puzzle pieces with the lowest numbers within each group because they are the same shape. This \emph{pruning} of pieces dramatically reduces the number of combinations to try; in the example, pruning away two pieces reduces the possible combinations from 5 to 2.


While the puzzle analogy is simple and intuitive, we need more considerations to apply to a mapspace search. We discuss these in-depth in the paper, but one example is that pruning pmappings involves a multi-faceted tradeoff. For example, pmappings often trade off resource usage (\eg on-chip buffer usage). Thus, the lowest energy pmapping for an Einsum may not lead to the lowest energy mapping overall if that pmapping reserves too much on-chip buffer, which limits how much data pmappings of other Einsums can buffer. To ensure we do not prune a potentially efficient pmapping, we generate \emph{all} pmappings in the Pareto-frontier in terms of criteria that include objectives (\eg energy and latency) and resource usage (\eg memory usage).


In summary, this paper makes the following contributions:

\begin{itemize}
    \item We enable grouping and joining pmappings by defining the \emph{compatibility criteria}, such that two pmappings may be joined (\ie satisfy data dependencies) if and only if they agree in terms of compatibility.

    \item We enable pruning within each pmapping group by capturing optimization objectives (\eg energy and latency) and hardware resource usage (\eg memory usage) as \emph{objective} and \emph{reservation criteria} respectively. Optimality of the final, overall mapping is ensured if we only prune pmappings worse in \emph{all} criteria (\eg high energy and latency while using a lot of memory). Section~\ref{sec:optimality_guarantee} provides a proof.

    \item We introduce the \ffmlong (\ffm), a fast algorithm to find optimal fused mappings in a large mapspace for a wide variety of workloads. \ffm constructs mappings one Einsum at a time, grouping and pruning before joining pmappings to keep the search space small. 

    \item We show that \ffm can find optimal mappings of workloads with many Einsums (experimentally verified up to 64 Einsums) despite the exponentially increasing mapspace size, demonstrating the effectiveness of our pruning.
    
    \item We demonstrate the value of \ffm by showing an accelerator that uses FFM's mappings to improve accelerator EDP by up to $1.8\times$ compared to TransFusion~\cite{transfusion}, a state-of-the-art Transformer accelerator with hand-optimized fusion.

    \item We evaluate \ffm against state-of-the-art automated mappers. \ffm's mappings enable $2\times$ lower accelerator EDP compared to prior mappers when given similar search time. Furthermore, prior works do not find optimal mappings even with $10,000\times$ more runtime than \ffm.


\end{itemize}



\section{Background}
We briefly discuss the Einstein summation (Einsum) notation~\cite{einsum, extensor, teaal, edge}, which precisely describes a wide range of tensor algebra computation steps, and the LoopTree notation~\cite{looptree,looptree_thesis}, which precisely describes a wide variety of fused (and unfused) multi-Einsum mappings.

\subsection{The Einsum Notation}
Computation steps in a tensor algebra workload (\eg a matrix multiplication or a convolution step in a DNN layer) can be described using the \emph{extended Einsum notation}~\cite{einsum, extensor, teaal, edge}. In the Einsum notation, input and output multi-dimensional data are \emph{tensors}, and the dimensions of the tensors are called \emph{ranks}~\cite{efficient_processing_of_dnn}. For example, in a matrix multiplication $Z = A \times B$, the matrices $A$, $B$, and $Z$ are the tensors, the dimensions $M$, $K$, $N$ are the rank names and the Einsum is $Z_{m,n} = A_{m,k} \times B_{k,n}$. The variables $m$, $n$, $k$, which index into ranks, are called \emph{rank variables}, and a summation is implied over rank variables that are present in the right-hand side of the equation but not in the left-hand side ($k$ in this example). By convention, rank variables are often the lowercase letters of the rank names.

\subsection{The LoopTree Notation}
We explain the \emph{LoopTree} notation~\cite{looptree,looptree_thesis} using an example workload with two matrix-vector multiplications:

\begin{align}
    \text{(Step 1) Einsum A: }& A_{nA} = I_{nI} \times WA_{nI,nA} \label{eq:einsumA} \\
    \text{(Step 2) Einsum B: }& B_{nB} = A_{nA} \times WB_{nA,nB} \label{eq:einsumB}
\end{align}

\begin{figure}
\centering
\includegraphics[width=0.95\linewidth]{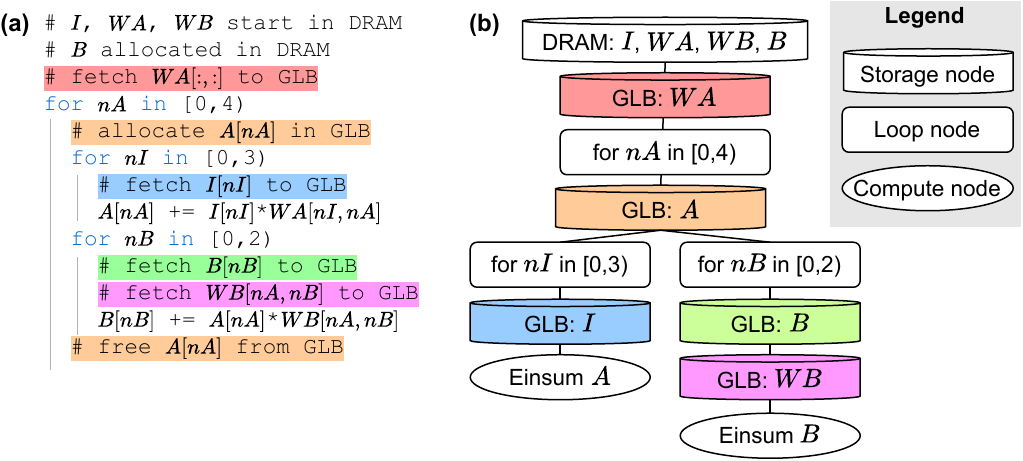}
\caption{(a) Pseudocode of a mapping for two matrix-vector multiplications and (b) the equivalent LoopTree mapping.}
\label{fig:mapping_examples}
\end{figure}

Fig.~\ref{fig:mapping_examples}(a) shows pseudocode for a fused mapping for these Einsums, and Fig.~\ref{fig:mapping_examples}(b) is the equivalent LoopTree mapping.

A LoopTree mapping consists of the following nodes:
\begin{itemize}
    \item \textbf{\emph{Loop nodes}} (rectangles) contain for-loops that iterate over rank variables in the workload. A loop may be shared by multiple fused Einsums (\ie an \emph{inter-Einsum} loop). For example, see the $nA$ loop, which represents co-iteration of multiple Einsum computations.
    \item \textbf{\emph{Storage nodes}} (cylinders) represent tensor tiles in memory. A storage node is annotated with the memory level (\eg DRAM or GLB) and the tensor stored in that memory level. Loops above the storage node determine the tile shape and iteration order. In our example, the placement of storage node ``GLB: $I$" indicates that the $NA$ rank is divided into four and the $NI$ rank into three to create twelve tiles of $I$. At each loop iteration, one such tile is fetched into GLB.
    \item \textbf{\emph{Compute nodes}} (ovals) indicate processing of one computation in the Einsum (\eg multiply-accumulate operations in this example).
\end{itemize}

In a LoopTree, important characteristics include the relative placement of nodes (\eg storage node over a loop node means tensor is reused across loop iterations) and splits (\eg loop above split means all Einsums below split interleave execution, storage nodes above splits means all Einsums under the split can use the same stored tensor tile).

\section{From Puzzle Analogy to Mapspace Search}
While the analogy in Section~\ref{sec:introduction} is intuitive, it oversimplifies the problem. In this section, we will (1) describe how \ffm constructs mappings from compatible pmappings, and (2) how to prune given the more complex tradeoffs of pmappings. Moreover, we will describe the \emph{criteria} with which we evaluate pmappings to ensure compatibility in (1) and optimality in (2).
\joel{In the first sentence the mention of compatibility and the "while ensuring optimality" phrase seems unnecessary, since you make that point in the next sentence.}
\michael{Updated.}

\subsection{FFM Constructs Mappings by Joining Compatible Pmappings}\label{sec:overview_examples}
\insightbox{\ffm constructs a full mapping by iteratively joining single-Einsum mappings, called \emph{partial mappings} or \emph{pmappings}. The \emph{compatibility criteria} capture whether pmappings are compatible and can be joined.}

\begin{figure}
    \centering
    \includegraphics[width=0.95\linewidth]{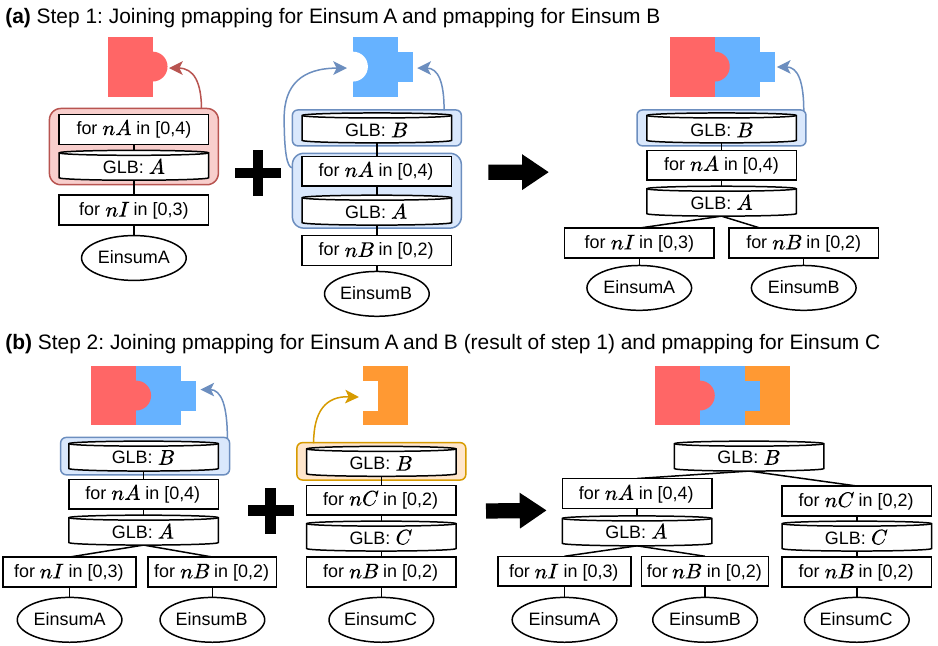}
    \caption{An example of creating a mapping in two joining steps. Weight tensors are omitted for clarity since they do not affect compatibility. (a) Joining pmappings for Einsums $A$ and $B$. (b) Joining the pmapping for Einsums $A,B$ and the pmapping for Einsum $C$. Storage nodes of shared tensors and the loops above them must match for pmappings to be compatible; these parts of the pmappings are highlighted in color and connected to the tabs/sockets analogy.}
    \label{fig:overview_example}
\end{figure}

\ffm constructs full mappings by iteratively joining compatible producer/consumer pmappings. In the following, we will describe the requirements for pmappings to be compatible and how pmappings are joined together. As an example, suppose we have as workload a cascade of three vector-matrix multiplications (insert after Eq.~\ref{eq:einsumB}, a new Einsum $C$, $C_{nC} = B_{nB} \times WC_{nB,nC}$), Fig.~\ref{fig:overview_example} shows a two-step process for joining three pmappings, one pmapping for each Einsum. We will walk through this example.

First, data dependencies between Einsums constrain which pmappings are compatible. These data dependencies occur when Einsums share tensors (\eg when an Einsum produces an intermediate that is consumed by another). For two pmappings to be compatible, they must agree on (1) the tile shape of shared tensors, (2) the memory level that keeps those tiles, and (3) the iteration order of those tiles (\ie the dataflow). \textbf{In a LoopTree, this means two mappings are compatible if they match in (only!) the storage node of shared tensors and the loops above these nodes.} For example, in Fig.~\ref{fig:overview_example}(a), a pmapping for Einsum $A$ is joined with a compatible pmapping for Einsum $B$. The shared tensor for Einsums $A$ and $B$ is only tensor $A$, and these pmappings are compatible because they have the same loop over $nA$ and storage node ``GLB: $A$" (see the highlighted portions of pmappings and their puzzle equivalents). Note that the ``GLB: $B$" is irrelevant to compatibility because it is not a shared tensor between Einsums $A$ and $B$.
\joel{I really don't see how the shared input fits in - the multiple consumer situation has to be handled by producer/consumer case and the multiple input situation is a degenerate form of that where there is a null Einsum $A_{out}=A_{in}$ that can be mapped. Is the multiple consumer case like a puzzle piece with two output tabs - which must be compatible in some way, \eg on is a subtile of the other.}
\michael{All of what you said is accurate. }
\joel{So I would remove the mention of "inputs" and just talk about producer/consumer relationships - you can make a note that input tensors and values used by multiple consumers are handled as well as generalized versions of the single producer/consumer case.}
\michael{Done.}

In \ffm, we capture all these characteristics that are relevant to compatibility as \emph{compatibility criteria}, which we will discuss in greater detail and precision in Section~\ref{sec:compatibility_criteria}.

When we join pmappings, we merge their compatible portions, making it one shared portion, and place a LoopTree split underneath that shared portion (see the result pmapping Fig.~\ref{fig:overview_example}(a)). Other portions of the pmapping (\eg the ``GLB: $B$" node) are inserted in their corresponding places in the result\footnote{There may be multiple equivalent ways to place these nodes. In a LoopTree, only the order of loop nodes and the relative positioning of storage nodes between the loop nodes matter. These are captured in the compatibility criteria definition.}. We can keep joining pmappings until we have the full mapping for the entire workload. For example, Fig.~\ref{fig:overview_example}(b) shows the result from Fig.~\ref{fig:overview_example}(a) being joined with a third pmapping for Einsum $C$. The shared tensor is now $B$, and the process from before can be applied in this step.

\renewcommand\fbox{\fcolorbox{red}{white}}

Finally, this process can construct a comprehensive mapspace: We can decompose any LoopTree mapping into individual pmappings, and conversely, there exist pmappings that join to create any given LoopTree mapping. Furthermore, there is a one-to-one correspondence between a LoopTree and a loop nest, so this procedure can be used to realize arbitrary loop nest mappings.

\subsection{Optimal Pruning under Complex Tradeoffs}
\insightbox{
  Pruning \emph{within} a group maintains optimality because the criteria used to prune within a group capture all tradeoffs between pmappings. Specifically, \emph{objective criteria} capture the contribution of a pmapping towards the overall objectives (\eg latency and energy), and \emph{reservation criteria} capture how much hardware resources (\eg memory capacity) are reserved by the pmapping.
}

Unlike the simple analogy in Section~\ref{sec:introduction}, in which pieces can be assigned a single numerical cost, choosing pmappings involves significantly more complex tradeoffs because pmappings compete for limited hardware resources. For example, one pmapping may have low energy but reserves most of the on-chip memory. Choosing this pmapping forces pmappings of other Einsums to use what little on-chip memory remains, which may reduce data reuse and increase energy overall. In general, choosing pmappings requires consideration of the efficiency of the pmapping in terms of our overall objective (\eg latency and energy) and the amount of hardware resources reserved. For each pmapping, we capture the former using \emph{objective criteria} and the latter using \emph{reservation criteria}.
 
\begin{figure}
    \centering
    \includegraphics[width=0.95\linewidth]{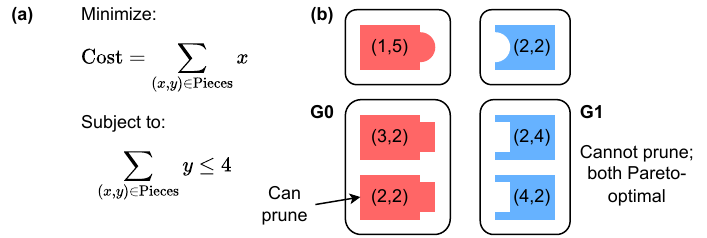}
    \caption{An example constrained optimization and (b) example puzzle pieces with two criteria (objective, reservation) to consider. Two groups of puzzle pieces, group G0 and group G1, are highlighted for discussion.}
    \label{fig:puzzle_tuple}
\end{figure}

To provide intuition, Fig.~\ref{fig:puzzle_tuple} shows a puzzle analogy in which each puzzle piece has an objective and a reservation of a resource (\eg memory capacity), and the goal is finding a combination that minimizes the sum of objectives while keeping the sum of reservations less than or equal to 4. This creates a new tradeoff where a piece may have a lower objective, but reserves too much of a resource.

We can prune while guaranteeing optimality of the final solution by pruning only pieces that are worse than another piece within that in both objective and reservation criteria (proof in Section~\ref{sec:optimality_guarantee}). To enable discussion, we will refer to a piece by its criteria tuple, (objective, reservation). Fig.~\ref{fig:puzzle_tuple}(b) shows an example in which group G0 contains two pieces with criteria (2,3) and (3,3). In G0, we can prune the piece with (3,3) because (2,3) is always better, with a lower objective and the same reservation. On the other hand, we cannot prune in group G1 because one has a lower objective while the other has a lower reservation. When there is a piece that is worse than another piece in all criteria, we refer to the former piece as being \emph{Pareto dominated}. After pruning all Pareto-dominated pieces, the remaining pieces in each group form the \emph{Pareto frontier}.

\subsection{Summary of Compatibility, Objective, and Reservation Criteria}\label{sec:criteria}


In summary, \ffm ensures joined pmappings respect data dependencies using the compatibility criteria, and evaluates the efficiency of pmappings using the objective and reservation criteria. Table~\ref{tab:taxonomy} organizes these criteria, points to sections that contain more detail, and shows examples and how the criteria are used in \ffm.

\newcommand{\tabcell}[2]{\multirow{#1}{*}{\begin{tabular}{c}#2\end{tabular}}}

\begin{table*}[t]
    \centering
    \setlength{\tabcolsep}{4.5pt}
    \begin{tabular}{lcccccc}
        \textbf{\thead{Criteria (Section)}} & \textbf{\thead[c]{Example}} & \textbf{\thead[c]{Puzzle Analogy}} & \textbf{\thead[c]{FFM Use}} & \textbf{\thead[c]{Overall Goal}} \\
    \toprule
        \makecell[l]{Objective (Section \ref{sec:criteria})}
                      & Pmapping Energy
                      & Piece cost
                      & \makecell[l]{Prune in group}
                      & Minimize \\
        \hline
        \makecell[l]{Compatibility (Section \ref{sec:compatibility_criteria})}
                      & \makecell[c]{Shape of shared tensor tile and which memory level \\ the exchange occurs and iteration order of tiles}
                      & Tab/socket shape
                      & \makecell[c]{Group}
                      & \makecell[c]{Satisfy data\\ dependencies}\\
        \hline
        \makecell[l]{Reservation (Section \ref{sec:reservation_criteria})}
                      & \makecell[c]{Tensor $X$ memory usage and\\for how long Tensor $X$ is alive}
                      & Piece weight
                      & \makecell[c]{Prune in group}
                      & \makecell[c]{Ensure below\\ resource limits}\\
    \toprule
    \end{tabular}
    \caption{A taxonomy of criteria that we use to compare and prune pmappings. Examples are illustrative, not full descriptions.}
    \label{tab:taxonomy}
\end{table*}

\section{Compatibility Criteria}\label{sec:compatibility_criteria}

A valid mapping must satisfy data dependencies (\eg data must be read after it is written and before it is freed). In this section, we discuss compatibility, which encompasses when and where pmappings exchange data. Moreover, we show how we encode compatibility as criteria (based on backing storage and loops above backing storage) to enable grouping and pruning within groups. 

\subsection{Formalizing Compatibility Criteria}
\insightbox{%
  Two pmappings are \emph{compatible} iff they agree on every
  shared tensor's (1)~tile shape, (2)~backing memory level, and (3)~tile production/consumption order (dataflow). \ffm's representation of compatibility criteria captures a comprehensive mapspace.
  \joel{What does this last sentence mean? Is it an assertion, consequence or unrelated to the insight? Or maybe the word "comprehensive" without qualification is throwing me.}
}
Pairs of pmappings must respect data dependencies to be compatible. Specifically, for every shared tensor, pmappings must agree on: (1) the shape of shared tensor tiles; (2) the memory level in which the tiles are stored, and (3) the order of production and consumption of the tiles (\ie the dataflow). These three attributes comprise the tensor's \emph{compatibility}.

Now, we formalize the compatibility criteria of a pmapping as the set $c_T$ for every shared tensor $T$, where $c_T$ is the 3-tuple
\begin{equation}
c_T = (\text{tile shape}, \text{backing storage}, \text{tile exchange order})
\end{equation}
In the LoopTree, the tile exchange order (dataflow) is the order of loop nodes above the storage node, the tile shapes are the bounds of those same loops, and the \emph{backing memory} (where tiles are exchanged between pmappings) is the highest storage node (outermost memory level) that stores tensor $T$.

Importantly, this representation encompasses a comprehensive mapspace. For example, \ffm allows any loop order as dataflow, which is much more flexible than many state-of-the-art designs, which only allow certain orders or impose a fixed dataflow~\cite{flat, fusemax, flashattention, flashattention2, defines, depfin}. Moreover, the backing memory choice encompasses the fusion design space. A pmapping can specify a backing memory to be DRAM (making it unfused), GLB (making it fused), or any other memory level. This lets us support a comprehensive fused space, including fusion at every memory level, rather than treating fusion as a separate case that must be considered. This choice is also more flexible than many state-of-the-art designs~\cite{flat, fusemax, transfusion, flashattention, flashattention2}.

\subsection{Compatibility for Grouping and Skipping Incompatible Joins}
\insightbox{Grouping based on compatibility lets us skip many incompatible joins and ensure pruning optimality because all pmappings in a group impose identical compatibility constraints on future pmappings.}

After generating the compatibility criteria for all pmappings, we group pmappings with the same compatibility to enable two optimizations: skipping joining incompatible pmappings and optimality-preserving pmapping pruning.

We reduce the time spent checking for pmapping compatibility by only joining pmappings from compatible groups. For example, instead of trying to join every pmapping for Einsum $A$ with every pmapping for Einsum $B$, we check the compatibility between groups for Einsum $A$ and groups for Einsum $B$. Then for any compatible group of $A$ and group of $B$, any pmapping in the group of $A$ is always compatible with any pmapping in the group of $B$. 

Moreover, grouping based on compatibility ensures that pruning is optimality-preserving because any two pmappings in the group are interchangeable. Specifically, just as we can trade one puzzle piece for another in the same group in Fig.~\ref{fig:puzzle_tuple}, we can trade any pmapping in a mapping with another pmapping from the same group. So, if pmapping $p_1$ is Pareto-better than pmapping $p_2$ in the same group, it is optimality-preserving to prune $p_2$ because any mapping containing $p_2$ could be improved by replacing $p_2$ with $p_1$.

\section{Reservation Criteria}\label{sec:reservation_criteria}
\insightbox{%
  Processing an Einsum requires reserving hardware resources. While joining pmappings, we track the current reservations at every possible lifetime, using them as pruning criteria.
  \joel{I don't know what "every possible lifetime" is. Do you mean "at every point of the tensor's lifetime"? I think these sections need a careful definition of lifetime and what is it that has a lifetime. }
}

Valid mappings must reserve system resources (\eg buffer capacity, number of processing elements) within available limits. Reservation limits represent a constraint, but also a pruning opportunity. All other attributes being equal, a pmapping that reserves more resources is a worse pmapping and a pruning candidate.

In this section, we discuss (1) calculating resource reservations from a \emph{full} mapping, (2) tracking the minimum amount of reservation information in a \emph{partial} mapping to still be able to compute the final reservations, and (3) pruning \emph{partial} mappings using reservation information.

\subsection{Calculating Reservations of Full Mappings}
\insightbox{
  Overall reservations can be computed by summing or taking the maximum of the reservations of each pmapping.
}

\begin{figure}
    \centering
    \includegraphics[width=0.9\linewidth]{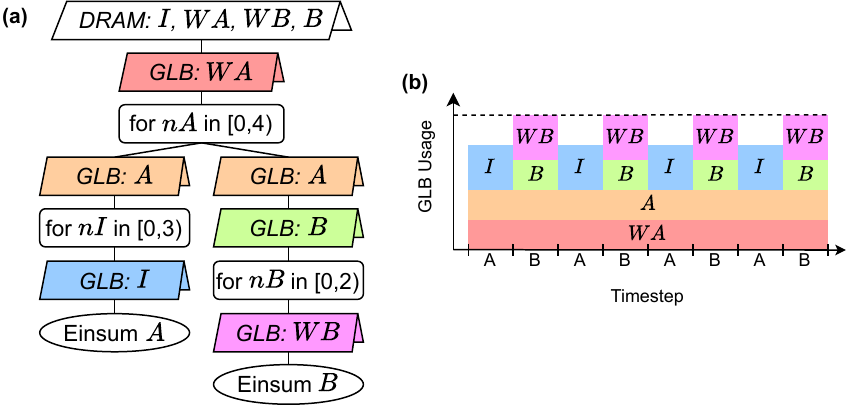}
    \caption{(a) ReservationTree for the mapping in Fig.~\ref{fig:mapping_examples}(b) replaces storage nodes (cylinders) with reservation nodes (placards), which show tensor tile lifetimes across branches explicitly. (b) Illustration of GLB usage. Color and label correspond to reservation nodes. Timesteps are marked with the Einsum being processed. The dashed line shows the maximum GLB usage.}
    \label{fig:resource_reservations}
\end{figure}

To illustrate how we calculate a mapping's reservations, we will calculate GLB usage for the mapping in Fig.~\ref{fig:resource_reservations}(a). We first convert the LoopTree to \emph{ReservationTree}; storage nodes immediately above splits are replaced with \emph{reservation nodes} in branches where the reservation is live (\eg ``$GLB: A$" in Fig.~\ref{fig:resource_reservations}(a)). Other storage nodes are replaced with reservation nodes at the same spot (\eg ``$GLB: WA$" in Fig.~\ref{fig:resource_reservations}(a)).

Fig.~\ref{fig:resource_reservations}(b) shows the resources reserved by the ReservationTree. Memory usage is not uniform during the processing because the sizes and lifetimes are different for different tensor tiles, but the maximum memory usage (dotted line) must be within the GLB capacity for the mapping to be valid.

All reservations within a branch are live at the same time, so we can calculate GLB usage for a branch as the sum of reservations in the branch (\eg Fig.~\ref{fig:resource_reservations}(a) has GLB reservations for $A$ and $I$, and Fig.~\ref{fig:resource_reservations}(b) shows $A$ and $I$ stacked in the same timestep). Different branches are active at different times, so we take a maximum of the GLB usage across branches (\eg in Fig.~\ref{fig:resource_reservations}(b), compare GLB usage for Einsum $A$ and $B$). This pattern generalizes: To calculate the total reservations of a ReservationTree, recursively sum reservations within a branch, and max reservations across branches.

\subsection{Reservation Lifetimes and Consolidating Reservations}
\begin{figure}
    \centering
    \includegraphics[width=0.9\linewidth]{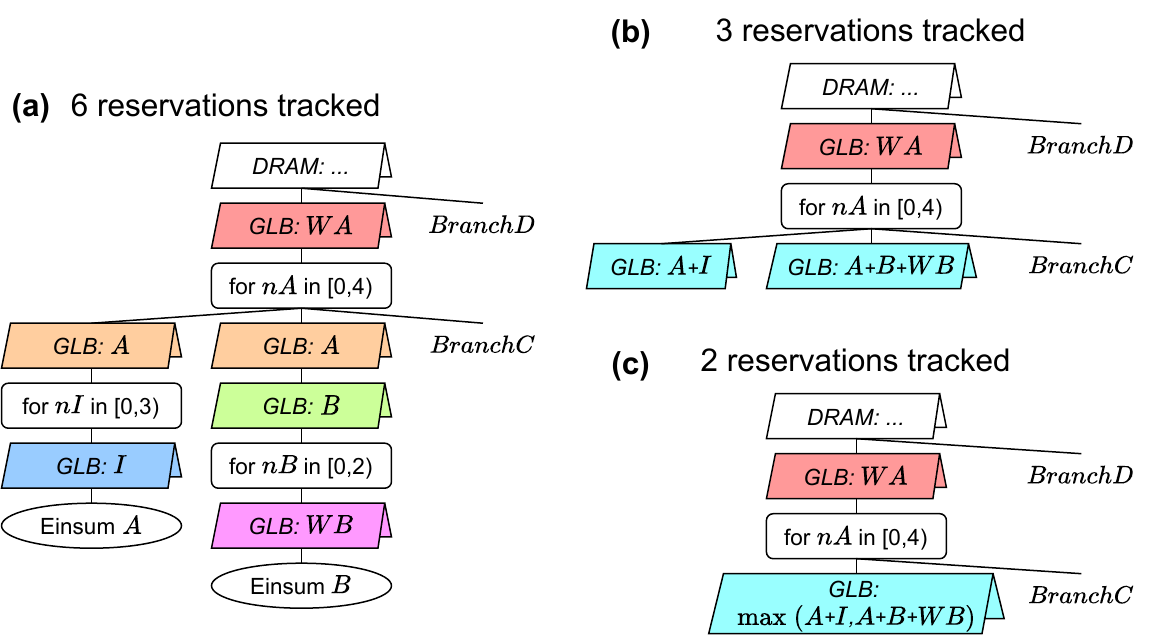}
    \caption{(a) The ReservationTree from Fig.~\ref{fig:resource_reservations} plus branches where future pmappings may be joined. (b) The ReservationTree after summing same-lifetime reservations. (c) The ReservationTree after maxing reservations across branches.}
    \label{fig:reservation_challenges}
\end{figure}

\insightbox{
  Each Einsum adds additional reservations. We can consolidate reservations with the same lifetime, letting us track only a limited number of reservations and avoid slowing down the mapper with too many pruning criteria.
}
The challenge in our method is that a pmapping alone does not tell us the lifetime of all reservations in the final mapping. Fig.~\ref{fig:reservation_challenges}(a) shows the same ReservationTree as in Fig.~\ref{fig:resource_reservations}, but now as a pmapping, and there are two places where a branch may be joined in the future: $BranchC$ and $BranchD$. A branch may be added under any loop or under the outermost DRAM storage node. In this example, we know the lifetime of the tensor $I$ tile in GLB is only within the timesteps when we process Einsum $A$. However, we do not yet know the lifetime of $WA$. If another Einsum branches off at $BranchC$, then $WA$ will be live while processing that Einsum. But if that Einsum branches off at $BranchD$, then $WA$ will only be live for Einsums $A$ and $B$, but not the newly joined Einsum.

These unknown lifetimes affect GLB usage, which we address by tracking each reservation. If $\hat{C}$ is GLB usage within $BranchC$, $\hat{D}$ is GLB usage within branch $D$, and $X$ is the memory usage of tensor $X$, then total memory usage is:
\begin{equation}
    \text{\textbf{Usage}}=\max(\hat{D},WA+\max(A+I,A+B+WB,\hat{C})) \label{eq:reservation_calculation}
\end{equation}
Note that although we do not yet know what $\hat{C}$ and $\hat{D}$ are, tracking the reservation of each tensor separately will allow us to compute the usage at the end.

This method requires tracking many reservations, but we can reduce this number using two observations: (1) there are only a few unique lifetimes (determined by branches), and (2) many lifetimes can be consolidated. First, notice that some lifetimes include neither $BranchC$ nor $BranchD$, others include $BranchC$ only, and others include all branches. When we compare pmappings, we can only compare reservations that have the same lifetime, and \emph{two reservations have the same lifetime if they are above the same branches.} This is true regardless of the contents of the branches, so it works even if we later add unknown pmappings to branches. So, any reservation between $BranchC$ and $BranchD$ splits could be compared to our $WA$ reservation because they would have identical lifetimes.

We can use the relatively few unique lifetimes to reduce the number of reservations to track. Fig.~\ref{fig:reservation_challenges}(b) shows that instead of tracking $A$, $I$, $B$, and $WB$ separately, we track the per-branch total:
\begin{equation}
    \text{\textbf{Sum same lifetimes:}}\ \hat{A} = A + I,\ \hat{B} = A + B + WB\label{eq:sum_same_lifetime_reservations}
\end{equation}
Then, we only have to track fewer values:
\begin{equation}
    \text{\textbf{Usage}}=\max(\hat{D},WA+\max(\hat{A}, \hat{B},\hat{C}))
\end{equation}

Moreover, branches that cannot have future pmappings joined to them (\ie branches on the left) can have their reservations consolidated into one value. Fig.~\ref{fig:reservation_challenges}(c) shows that the branches for Einsums $A$ and $B$ can be consolidated:
\begin{equation}
    \text{\textbf{Max across branches:}}\ \widehat{AB} = \max(\hat{A}, \hat{B}) \label{eq:max_complete_reservations}
\end{equation}
Then, we reduce the number of values to track even more:
\begin{equation}
    \text{\textbf{Usage}}=\max(\hat{D},WA+\max(\widehat{AB}, \hat{C})) \label{eq:final_usage}
\end{equation}

A crucial benefit of consolidation it means \textbf{the number of reservations to track does not depend on the number of Einsums.} If the next split occurs $\leq N$ loops down, then we only need to track up to $2N+1$ reservations.\footnote{The factor 2 comes from the fact that two lifetimes must be tracked for every potential split: reservations with a lifetime that end before future Einsums and reservations with a lifetime that continues to future Einsums. The constant 1 is for reservations below the bottommost split, which never live through future Einsums.} The split point is limited by the number of loops above the current split in each pmapping, which is limited by the number of shared ranks between the most recent and next Einsum. This number is generally small (\eg 4 for transformers).

\subsection{Pruning Pmappings using Reservations} \label{sec:consolidating}
To prune using reservations, we need to consider both the size of each reservation and the time period for which each reservation is alive. First, note that reservations can be compared for two pmappings which have the same expression for resource usage (\eg the same expression as Eq.~\ref{eq:final_usage} just with different values for $WA$ and $\widehat{AB}$). Equivalently, two pmappings can be compared if they have the same ReservationTree form after consolidation (\eg the tree in Fig.~\ref{fig:reservation_challenges}(c)). Note that a ReservationTree after consolidation reduces to a subtree that contains the same information as the compatibility criteria. Therefore, any two pmappings with the same compatibility criteria can have their reservations compared.

To compare the reservation of two pmappings and ensure pruning preserves optimality, we consider all reservations and check for Pareto dominance (\ie a pmapping has worse reservations if and only if its reservation is larger across all lifetimes for all resources).

\section{\ffmlong (\ffm)}\label{sec:ffm_overview}
\insightbox{%
  FFM works in two phases: (1)~\emph{single-Einsum exploration} generates Pareto-optimal pmappings for each Einsum and (2)~\emph{iterative group--prune--join} processes one Einsum at a time. Iterative pruning limits the active Pareto-frontier size, which reduces mapper runtime and memory overhead.
}

\begin{figure}
    \centering
    \includegraphics[width=0.77\linewidth]{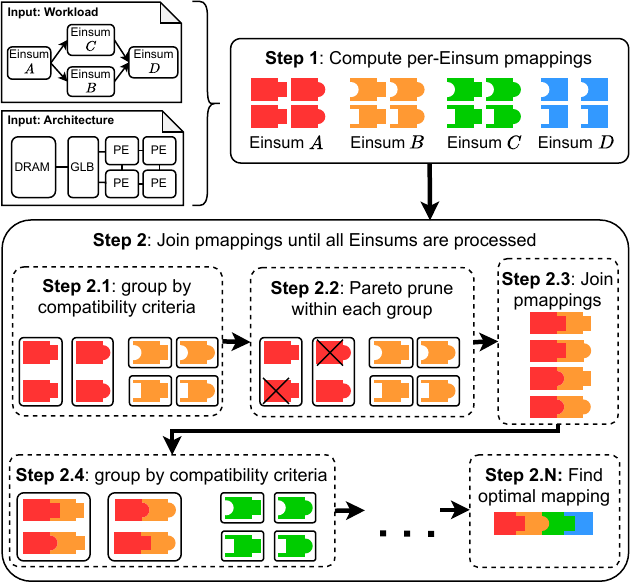}
    \caption{Overview of the \ffmlong (\ffm).}
    \label{fig:fastfusion_overview}
\end{figure}

The \ffmlong (\emph{\ffm}) implements the ideas in the previous sections to quickly find optimal mappings. Fig.~\ref{fig:fastfusion_overview} shows \ffm, which has two main steps. First, \ffm explores the space of pmappings for each Einsum. Next, starting with the first Einsum, it repeatedly groups, prunes, and joins pmappings until all Einsums are joined.

\subsection{Fast Single-Einsum Pmapping Generation}
We generate all Pareto-optimal pmappings for every Einsum, exploring dataflow, tile shape, storage node placement, and spatial allocation. To explore all fusion options, the pmapping exploration considers all backing storage choices.\footnote{While we show one processing order for the Einsums, there may be more than one data-dependency-satisfying processing order of Einsums. FFM can explore these orders by joining in all data-dependency-respecting orders of Einsums. Pmappings are the same for any order and are only computed once.}

While an exhaustive exploration of all pmappings would be slow~\cite{timeloop}, we use TCM~\cite{turbocharged}, which can quickly generate Pareto-optimal pmappings by pruning the search space by more than 30 orders of magnitude.

\subsection{Group, Prune, and Join Pmappings}
We explain this section with an example, then extend it to the full algorithm. Assume we have the pmappings for Einsum $A$ and we'd like to join them with the pmappings for Einsum $B$. We would perform the following steps:

\begin{itemize}
    \item \textbf{Group} the pmappings for Einsum $A$ based on compatibility criteria. Do the same for Einsum $B$.
    \item \textbf{Prune} dominated pmappings in each group by eliminating pmappings not on the Pareto frontier.
    \item \textbf{Join} compatible pmappings for each group combination. For every valid combination of groups, we generate every combination of pmappings.
\end{itemize}

\tanner{Note for camera-ready: Prune $\rightarrow$ extract frontier.}
\tanner{Note for camera-ready: Skipping incompatible joins $\rightarrow$ only join compatible pmappings}

After performing these steps, we will have pmappings that map both Einsums $A$ and $B$. We repeat the process, adding pmappings for one Einsum at a time until we have processed all Einsums.

We use several optimizations to speed up this process. First, a given pmapping for one Einsum will likely only be compatible with a fraction of pmappings for another Einsum, so we \textbf{skip incompatible joins} by pairing groups of pmappings first based on their compatibility, then joining pmappings only between compatible groups (described in Section~\ref{sec:compatibility_criteria}).

Next, each Einsum adds more reservations to track, so after each join step we \textbf{consolidate reservations} (described in Section~\ref{sec:reservation_criteria}) to keep pruning effectively.

\subsection{Why Per-Einsum Runtime Does Not Increase With More Einsums}\label{sec:linear_time}
The astute reader may wonder, if the number of mapping choices increases exponentially with more Einsums, how does FFM stay fast? In practice we find that the runtime scales approximately linearly with the number of Einsums for two reasons.

First, the runtime of generating pmappings is linear in the number of Einsums because it is performed independently for each Einsum. This runtime generally dominates the overall FFM runtime.

Joining runtime does not increase because we keep the number of pmappings bounded in each joining step. To do so, we note that we keep pmappings that are Pareto-optimal in terms of compatibility, reservations, and objective metrics. There are finite compatibility choices (because they depend on the loops of only the most recent Einsum) and finite reservations (because we perform consolidation), so the number of dimensions in the Pareto space is bounded. We only need to ensure that the density of the Pareto-optimal space remains bounded.

To keep the Pareto-optimal space from becoming arbitrarily dense, we use $\epsilon$-pruning~\cite{epsilon_pruning} to perform a ``dirty" join, yielding mappings within $(1+\epsilon)\times$ optimal in the chosen objective. $\epsilon$-pruning coarsens the Pareto frontier to bound the number of points~\cite{epsilon_pruning}. However, it may yield mappings up to $(1+\epsilon)\times$ optimal in each objective, so we use two strategies to preserve optimality:

\begin{itemize}
    \item \emph{Resource Reservations}: If $\epsilon$-pruning reservations yields a valid mapping, then it is known to be optimal because we only care if reservations are valid (\eg if a mapping has the lowest energy and has valid resource usage, we don't care whether the resource usage could be $\epsilon$ lower). If $\epsilon$-pruning yields an invalid mapping, we retry with a smaller $\epsilon$.
    \item \emph{Objectives}: We first perform a dirty search with $\epsilon$-pruning, yielding a near-optimal mapping with objectives \(\leq(1+\epsilon)\times\) optimal. Then we perform a second clean search and, after each joining step, prune pmappings worse than our near-optimal mapping. Most pmappings are significantly worse than optimal~\cite{timeloop}, so this pruning removes most pmappings.
\end{itemize}

In cases where there is one objective metric (\eg energy or latency) and dirty resource reservation pruning succeeds for $\epsilon>0$, the number of mappings stays bounded and we can guarantee runtime is linear with number of Einsums.

In theory, runtime may be exponential with the number of Einsums if dirty resource reservation pruning fails or if we have multiple objective metrics (\eg optimize both energy and latency $\rightarrow$ runtime may increase with the number of mappings on the energy-latency Pareto). In practice, however, we have not seen cases that scale superlinearly. Dirty resource reservation pruning nearly always succeeds on our first $\epsilon$ guess of $0.2$, and we have never seen it fail for $\epsilon\leq0.02$. Meanwhile, multi-objective optimization has been fast for all tested workloads (\eg all experiments in this paper use \ffm returning the full energy-latency-optimal Pareto, and Fig.~\ref{fig:runtime_scaling} shows that runtime does not increase).

\subsection{FFM finds Optimal Mappings \label{sec:optimality_guarantee}}
FFM finds optimal mappings in our mapspace because it \textbf{(1)} explores every non-pruned mapping and \textbf{(2)} ensures all pruning is optimality-preserving. Here, we prove both of these statements.
\joel{I think you need to reiterate that it is optimal for the mapspace we've defined.}
\michael{Updated.}

To satisfy \textbf{(1)}, we explore every pmapping for every Einsum with TCM~\cite{turbocharged}, which can fully explore mapspaces. Then, when joining, we enumerate every valid combination of pmappings, thus yielding every possible mapping choice.

To prove \textbf{(2)}, we show that if pmapping $X$ is pruned, then for any mapping containing $X$, there exists another mapping that is both \textbf{(A)} valid, data-dependency-satisfying, and same-or-better in every objective metric, and \textbf{(B)} not pruned.

Proof of \textbf{(A)}: $X$ is pruned if and only if there exists a pmapping $Y$ that dominates $X$, meaning $X$ and $Y$ have identical compatibility and $Y \leq X$ in every resource usage and metric. Therefore, any valid mapping that contains $X$ could use $Y$ instead, and the result would be valid ($Y\leq X$ in usage of every resource), satisfy data dependencies ($Y$ and $X$ have identical compatibility), and be better ($Y\leq X$ in objective metrics).

Proof of \textbf{(B)}: Define the set $S$ of pmappings that dominate $X$. If $S_0=\varnothing$, then $X$ is non-dominated and is not pruned. Otherwise, prove by induction. Assume true for $|S|<n$. For $|S|=n$, pick $Y\in S$ and define $S'$ as the set of pmappings that dominate $Y$. Any pmapping that dominates $Y$ also dominates $X$, and $Y$ does not dominate itself, so $|S'|< |S| = n$. By our assumption, either $Y$ is non-dominated or $S'$ contains a non-dominated pmapping.


\section{Comparison to Prior Mappers}\label{sec:results}

\insightbox{FFM quickly finds optimal mappings, while prior mappers fail to do so, even in orders-of-magnitude more runtime. This is because an optimal mapping must balance many tradeoffs across memory levels and Einsums.}


\subsection{Accelerator Architecture \& Validation}
We evaluate using an architecture based on TPUv4i~\cite{tpuv4i}, which has a 128 MiB global buffer and four cores. Each core has a 4 MiB local buffer and a 128$\times$128 PE-array that supports 8-bit multiply-accumulate (MAC) operations, operating with a weight-stationary dataflow. The architecture operates at 1.05 GHz, and connects to DRAM with 614 GB/s bandwidth.

\textbf{Hardware Model:} The hardware model is based on the validated LoopTree~\cite{looptree} model and component energies are modeled with HWComponents~\cite{cimloop, accelergy}. The energy and throughput of all components are taken directly from the TPUv4i~\cite{tpuv4i} paper. We validate our model against LoopTree, Timeloop, and CiMLoop~\cite{looptree,timeloop,cimloop}, and it yields identical results. Additionally, FFM’s open-source release includes the full CiMLoop suite, with models of five fabricated~\cite{jia,sinangil,wan,wan_ii,wang,wang_ii,colonnade} and four simulated~\cite{albireo,raella,lightning,isaac} chips, validated to match energy, area, and latency within 10\% of published data under a range of workloads and operating conditions (e.g., varied supply voltage).

\subsection{Baselines}

We compare our mapper to three baselines implemented based on prior mappers\footnote{Each paper's implementation supports a different mapspace, so we create mappers for this mapspace based on their approach.}.

\textbf{SET}~\cite{set}, uses simulated annealing to explore storage node placements, shared loops for each Einsum, and intra-Einsum by randomly changing choices and accepting the new one probabilistically. Cooling rate and initial temperature are taken from the SET implementation~\cite{set_implementation}.
 
\textbf{TileFlow}~\cite{tileflow} uses a genetic algorithm to pick storage node placements, shared loops for each Einsum, and intra-Einsum mapping choices. Each population has 10 randomly-initialized mappings (2000 total across 192 threads)\footnote{We evaluated larger population sizes, but those converged more slowly.}. We set a crossover rate of $0.7$ and a mutation rate of $0.2$, which are the best-converging options found for eight matrix multiplications.

\textbf{Timeloop}~\cite{timeloop} randomly samples a pmapping for each Einsum.

\subsection{Methodology}

For \textbf{\ffm} and all baselines, we report the energy-delay-product (EDP) of the mappings found and the runtime of each mapper. Runtime reported for \textbf{\ffm} is the wall-clock time, which includes the exploration of single-Einsum pmappings and the process of joining and pruning pmappings. 

Baselines would take CPU-years to run directly, so we generously estimate runtime (a lower bound of actual runtime). To estimate baseline runtime, we run the baselines, but replace their mapping evaluation calls with queries to the cached results from \textbf{\ffm's} pmapping explorations. We model baseline runtime by multiplying the number of queried pmappings by the amount of time it takes to evaluate a pmapping.

This estimated runtime is a generous underestimate for multiple reasons. First, we ignore the time taken by the baselines to join pmappings and to select new pmappings (\eg for \textbf{TileFlow}, we do not include the time required to perform crossover and mutation). Second, we let the baselines benefit from FFM's batched pmapping generator calls (asking the pmapping generator for all Pareto-optimal pmappings, rather than one pmapping at a time), which speeds up the pmapping generator by orders of magnitude~\cite{turbocharged}. Third, we only give baselines compatibility-valid pmappings, dramatically reducing their required search space. Baselines only initialize mappings with compatible choices, and when \textbf{TileFlow} or \textbf{SET} transform one pmapping, other pmapping(s) are also transformed into compatible equivalents.

Experiments ran on 2 Intel Xeon Gold 6252 and 384GB of RAM.

\subsection{\ffm Quickly Finds Optimal. Baselines Cannot Even With $10,000\times$ More Time}\label{sec:results:comparison}
We compare to the baselines mapping to TPU-v4i~\cite{tpuv4i} the GPT-3 6.7B Transformer model~\cite{gpt3} with batch size 64 and 4096 tokens. The batch size and token count represent a common scenario in which shared tensors do not fit in on-chip memory without tiling them, making it nontrivial to find a good fused mapping.\footnote{If shared tensors fit on-chip, one may fuse by keeping them on-chip without tiling.} \textbf{\ffm} finished after 1.5 CPU hours. We terminate baselines after $10,000\times$ the time taken by \textbf{\ffm}.

\begin{figure}
    \centering
    \includegraphics[width=0.95\linewidth]{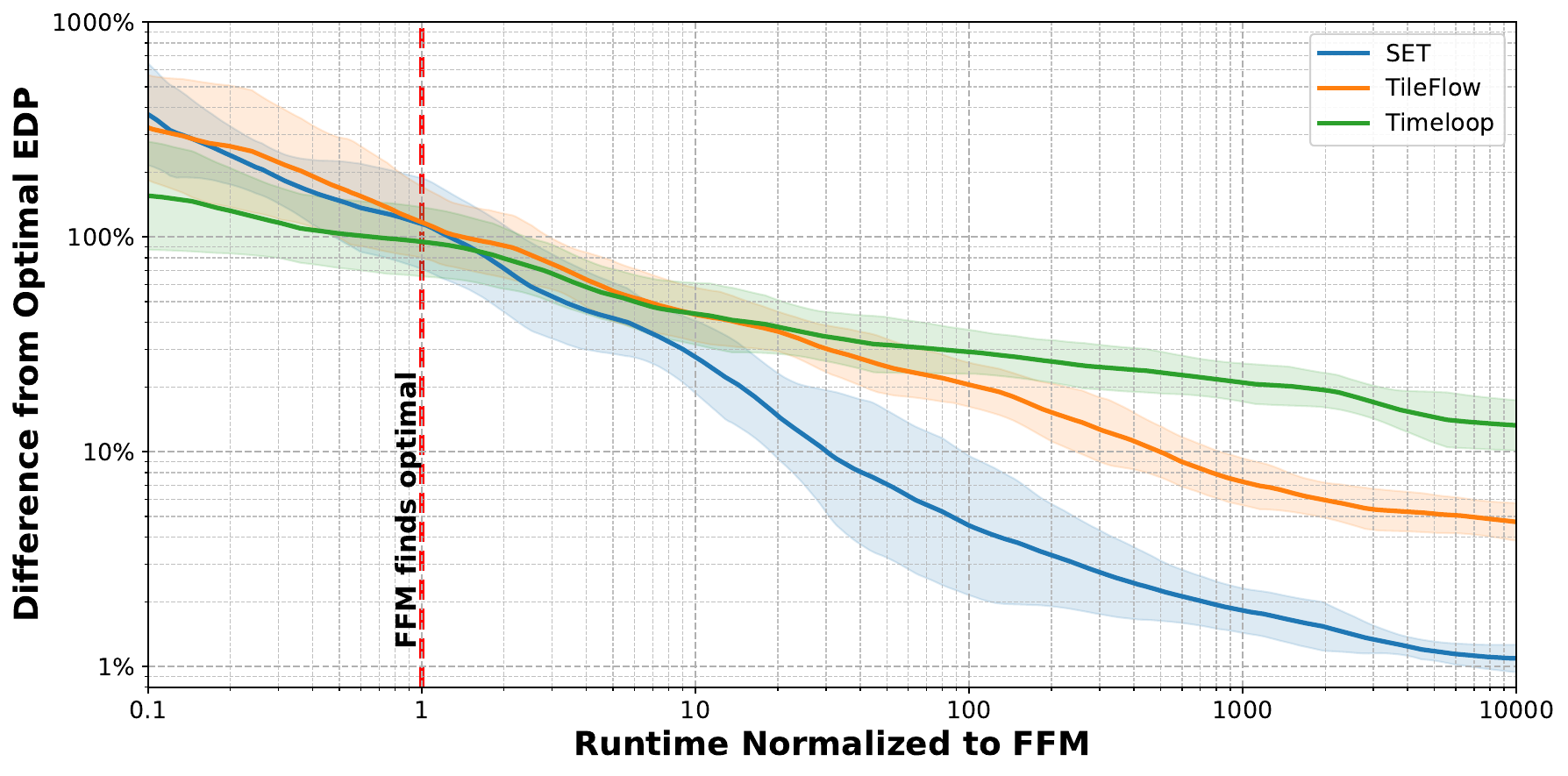}
    \caption{Convergence speeds of baselines relative to \textbf{\ffm}. \textbf{\ffm} finds the optimal mapping while baselines are still $100\%$ away from optimal EDP. Baselines are stochastic, so we show the average and $\pm1$ standard deviation range of ten runs. Baselines are $\geq1\%$ away from optimal EDP even at $10,000\times$ \textbf{\ffm}'s runtime. \textbf{\ffm} ran in $\approx1.5$ CPU hours.}
    \label{fig:baseline_convergence}
\end{figure}

Fig.~\ref{fig:baseline_convergence} compares how fast the baselines converge to the optimal EDP relative to \textbf{\ffm}. The y-axis shows the percent difference from optimal EDP. We can see that by the time \textbf{\ffm} completes, the baselines are approximately $100\%$ away from optimal ($2\times$ optimal). Given additional time, baselines converge slowly. \textbf{SET}, the best-performing mapper, is $30\%$ away from optimal after $10\times$ our runtime and $6\%$ off optimal after $100\times$ our runtime. To yield a mapping within $1\%$ of optimal, \textbf{SET} requires $15,000$ CPU hours (1.7 CPU years), making search infeasible for most design space explorations. 

\subsection{\ffm is Fast Even with Many Einsums} \label{section:runtime_comparison}
To compare the speed of our mapper against the baselines as a function of the number of Einsums, we run the mappers on chains of matrix multiplications (the output of one matrix multiplication is the input to the next and so on), and vary the number of matrix multiplications. Each matrix multiplication has $M=$ 8,192; and we create a variety of tensor shapes by using the following pattern for $(N;K)$: (16,384; 16,384) $\rightarrow$ (4,096; 16,384) $\rightarrow$ (4,096; 4,096) $\rightarrow$ (16,384; 4,096) $\rightarrow$ repeat. 

We time out baselines after $1000\times$ \textbf{\ffm}'s runtime. Like in the previous experiment, we average the results of ten runs for the baselines. When five or fewer of the ten runs time out, we generously assume the timed-out runs finish in  \textbf{\ffm}'s runtime. When six or more time out, we consider the baseline to have timed out.

\begin{figure}
    \centering
    \includegraphics[width=0.95\linewidth]{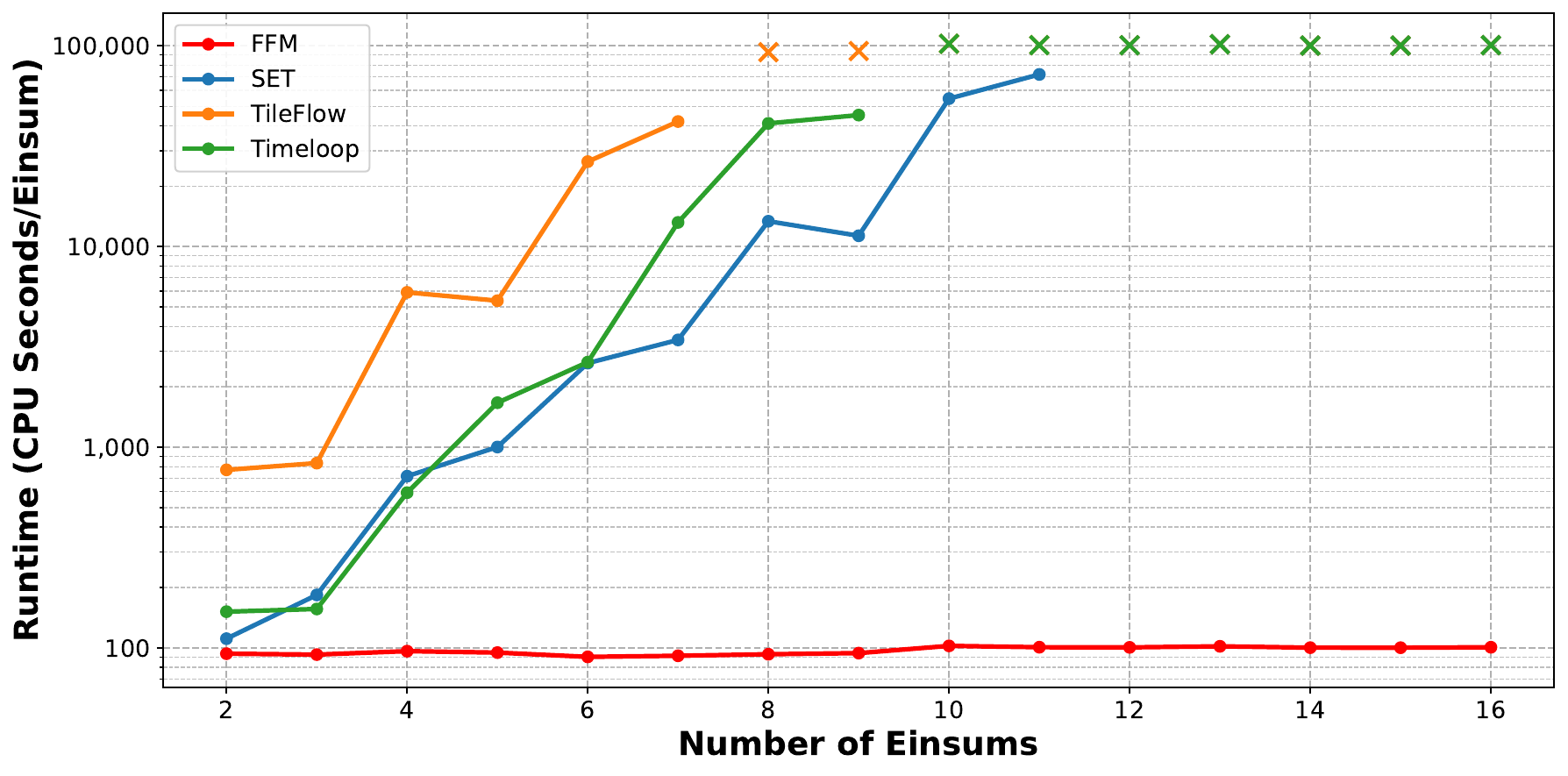}
    \caption{\ffm is fast and per-Einsum runtime remains flat even with more Einsums. The plot shows runtime for \ffm to find optimal mappings and baselines to find a mapping within $5\%$ of optimal. $\times$ denotes baseline timeouts (\(>1000\times\) \textbf{\ffm} runtime). Baseline runtime grows exponentially with more Einsums.}
    \label{fig:runtime_scaling}
\end{figure}

Fig.~\ref{fig:runtime_scaling} shows the time taken by \textbf{\ffm} to find an optimal mapping and the baselines to find a mapping within $5\%$ of optimal. \textbf{\ffm} quickly finds optimal mappings for all numbers of Einsums, while the runtime of the baselines rapidly increases with more Einsums. 

In fact, \textbf{\ffm}'s per-Einsum runtime is approximately constant, meaning that runtime scales linearly with more Einsums, despite the exponential mapspace size. Baseline runtime increases exponentially with more Einsums and time out. Here, we used a near-optimal threshold of $5\%$ to get more baseline data, since thresholds closer to optimal led to baselines timing out in as few as two Einsums.

Theoretical guarantees of linear time are nuanced, so we refer the reader to Section~\ref{sec:linear_time} for a deeper discussion.

\subsection{\ffm Finds Efficient Tradeoffs Across Einsums and Memory Levels}\label{sec:many_tradeoffs}
\begin{figure}
    \centering
    \includegraphics[width=0.95\linewidth]{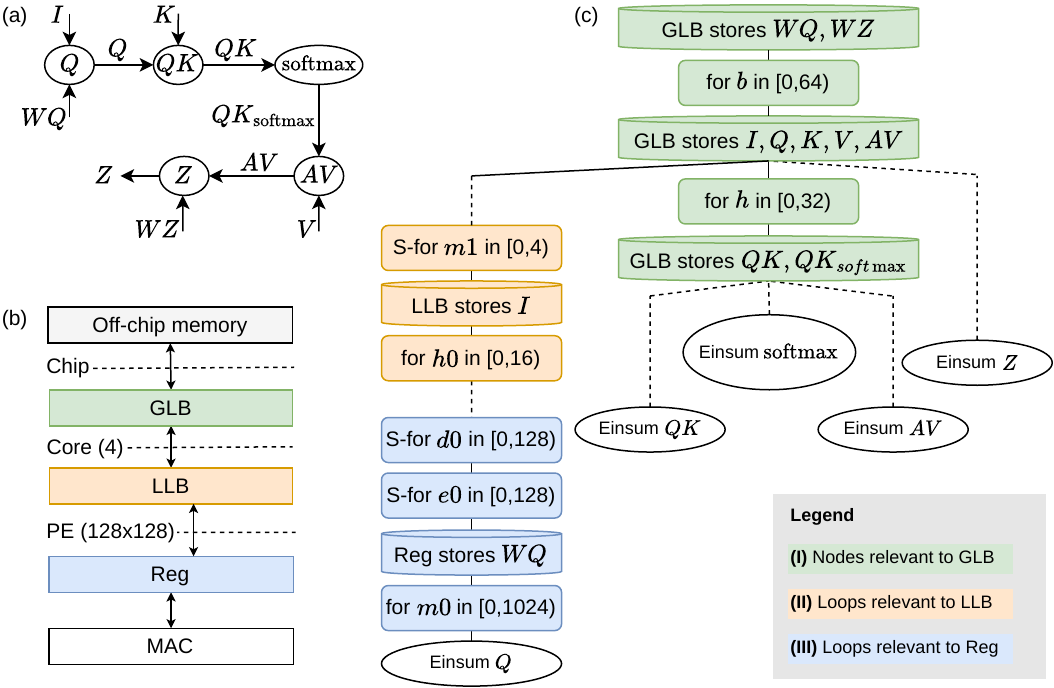}
    \caption{\ffm balances tradeoffs across Einsums and across the memory hierarchy. (a) A subset of the GPT-3 workload we will focus on. Ovals are Einsums and arrows are tensors. (b) The memory hierarchy of the TPU-v4i architecture. There are four cores, and each core has a 128$\times$128 PE array. (c) The relevant subset of the \ffm mapping. Dashed lines indicate omitted nodes. Certain nodes are highlighted for discussion.}
    \joel{Is it really TPU-v4i?}
    \michael{This is our model of it, yes. Are you suggesting we hedge and say ``TPU-v4i-like"?}
    \label{fig:optimal_mapping}
\end{figure}

Because of how we group, join, and prune, \ffm produces mappings that achieve efficient tradeoffs across Einsums in the workload and memory levels in the architecture. In this section, we highlight some of these tradeoffs. For this discussion, Fig.~\ref{fig:optimal_mapping}(a) shows a subset of GPT-3 Einsums that we discuss here. Moreover, Fig.~\ref{fig:optimal_mapping}(b) shows the memory levels of the TPUv4i accelerator. Finally, Fig.~\ref{fig:optimal_mapping}(c) shows the \ffm mapping with most of its nodes omitted to highlight the ones we will discuss. We make two observations.

First, we show how \ffm leverages per-Einsum reuse opportunities to efficiently utilize resources shared across Einsums. Nodes (I) in Fig.~\ref{fig:optimal_mapping}(c) exemplify this fact. Notice that tensors from various Einsums are kept in the on-chip GLB. However, there is not enough GLB capacity to keep the entirety of all these tensors. Instead, as shown, some tensors are tiled, and different tensors are tiled differently in a way that leverages particular reuse opportunities. For example, because different elements in a batch reuse weights, weights are kept in entirety in the GLB to be reused across a batch. In contrast, tensors such as $I$, $Q$, and $K$ (which do not have reuse opportunity across batch) are tiled, and only one element in a batch is in GLB at a time. Similarly, tensors $QK$ and $QK_\text{softmax}$ are tiled in the heads rank since there is no reuse across heads of these tensors.

Second, \ffm finds efficient tradeoffs across the memory hierarchy.
Nodes (II) and (III) in Fig.~\ref{fig:optimal_mapping}(c) show loops over $m1$ and $m0$, which iterate over different tokens. The loop ``S-for $m1$" indices parallelization over LLBs in different cores. The loop ``for $m0$" indicates iterations over tokens at the PE level, which reuses the weights kept in PE register Reg. This full utilization of LLBs (all 4 are used) and efficient reuse in Reg (1024$\times$ reuse of $WQ$) is possible because the tiles of $I$ and $Q$ have enough tokens. If a smaller tile is kept on-chip in the GLB (a decision made by nodes (I)), there would not have been as many tokens to parallelize or to reuse weights with.


\section{An Optimized Accelerator Built Using \ffm}\label{sec:result:comprehensive_mapspace}
\insightbox{We propose a transformer accelerator with novel mappings, generated using \ffm from a comprehensive mapspace, which achieves lower energy-delay-product (up to 1.8$\times$ for GPT-3) compared to TransFusion, a state-of-the-art baseline with hand-optimized fusion.}

We design an accelerator for edge inference using \ffm to determine the optimal mapping, which includes deciding which Einsums to fuse. The accelerator, which we refer to as \textbf{\ffm}, is organized like the TPU-v4i~\cite{tpuv4i} and sized to achieve $\sim33$ TOPS, comparable to current edge accelerators~\cite{copilot_plus_microsoft}. It consists of a 5 MB global buffer, a 128$\times$128 matrix-multiplication unit, and a vector unit with 256 compute units, which handles softmax and normalization. The accelerator is connected to an off-chip LPDDR4~\cite{accelergy,cimloop,hwcomponents} memory. We estimate GlobalBuffer energy using the Accelergy~\cite{cimloop,accelergy} interface to CACTI~\cite{cacti7}. Tab.~\ref{tab:energy_table} shows energy and bandwidth details

As baseline, we compare against \textbf{TransFusion}~\cite{transfusion}, a state-of-the-art Transformer accelerator. Importantly, \textbf{TransFusion} always fuses all Einsums except $K$ and $V$. We also implement the dataflow in~\cite{transfusion}, which produces and immediately consumes intermediate tensor tiles one at a time, refetching weights, $K$, and $V$ as needed (\ie an intermediate-tensor-stationary dataflow). Finally, we provision the baseline to match \textbf{\ffm}'s number of compute units and on-chip buffer size. We evaluate both accelerators on GPT-3 6.7B~\cite{gpt3} with a batch size of 1.
\joel{Because we changed the provisioning of the Transfusion accelerator, why do we expect that the mappings they used is still appropriate?}
\michael{Updated. In short, we follow their method for optimizing for a given provisioning. Our provisioning is also between their ``edge" and ``cloud" configurations.}

\begin{table}[b]
    \centering
    \begin{tabular}{l l l l}
        \textbf{Component} & \textbf{Type} & \textbf{Energy} & \textbf{Bandwidth} \\
        \toprule
        Off-chip memory & LPDDR4   & 8 pJ/b    & 30 GB/s  \\
        Global buffer   & SRAM     & 0.2 pJ/b  & 512 GB/s \\
        MAC             & int8-MAC & 0.08 pJ/b & 1 GHz* \\
        \bottomrule
    \end{tabular}
    \caption{Energy and bandwidth (*frequency for compute units) of accelerator components.}
    \label{tab:energy_table}
\end{table}

\begin{figure}[b]
    \centering
    \includegraphics[width=0.91\linewidth]{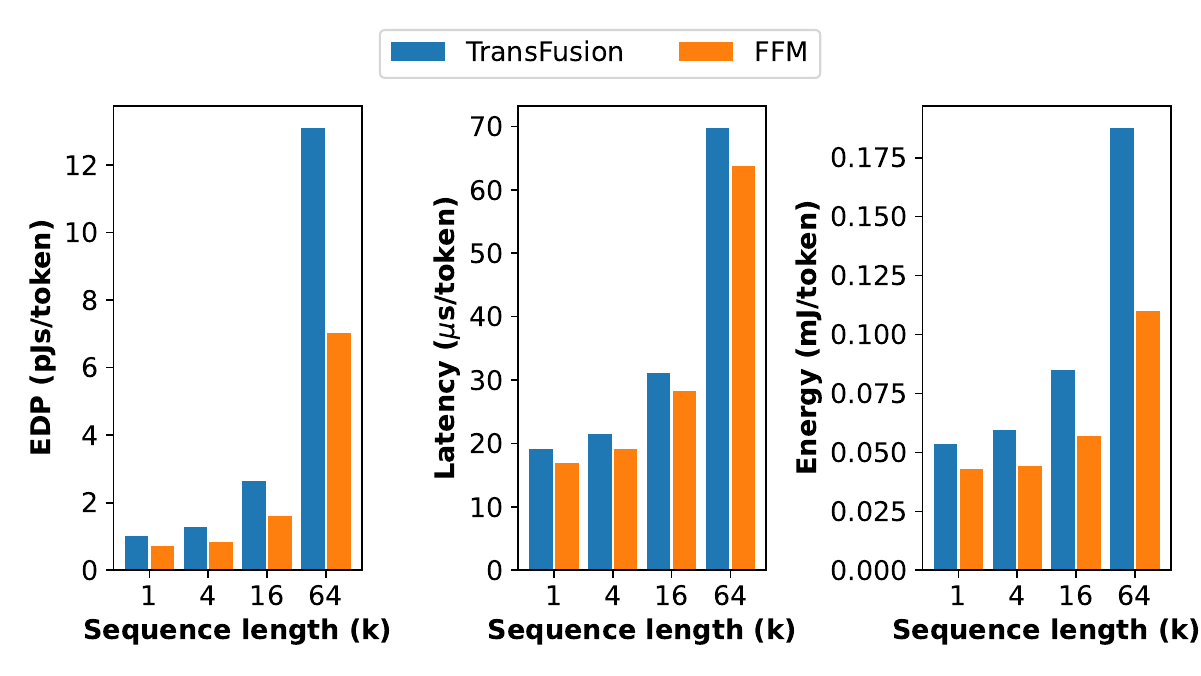}
    \caption{\ffm reduces EDP, latency, and energy per token generated compared to TransFusion~\cite{transfusion} for GPT-3.}
    \label{fig:comparison_summary}
\end{figure}

\begin{figure}
    \centering
    \includegraphics[width=0.95\linewidth]{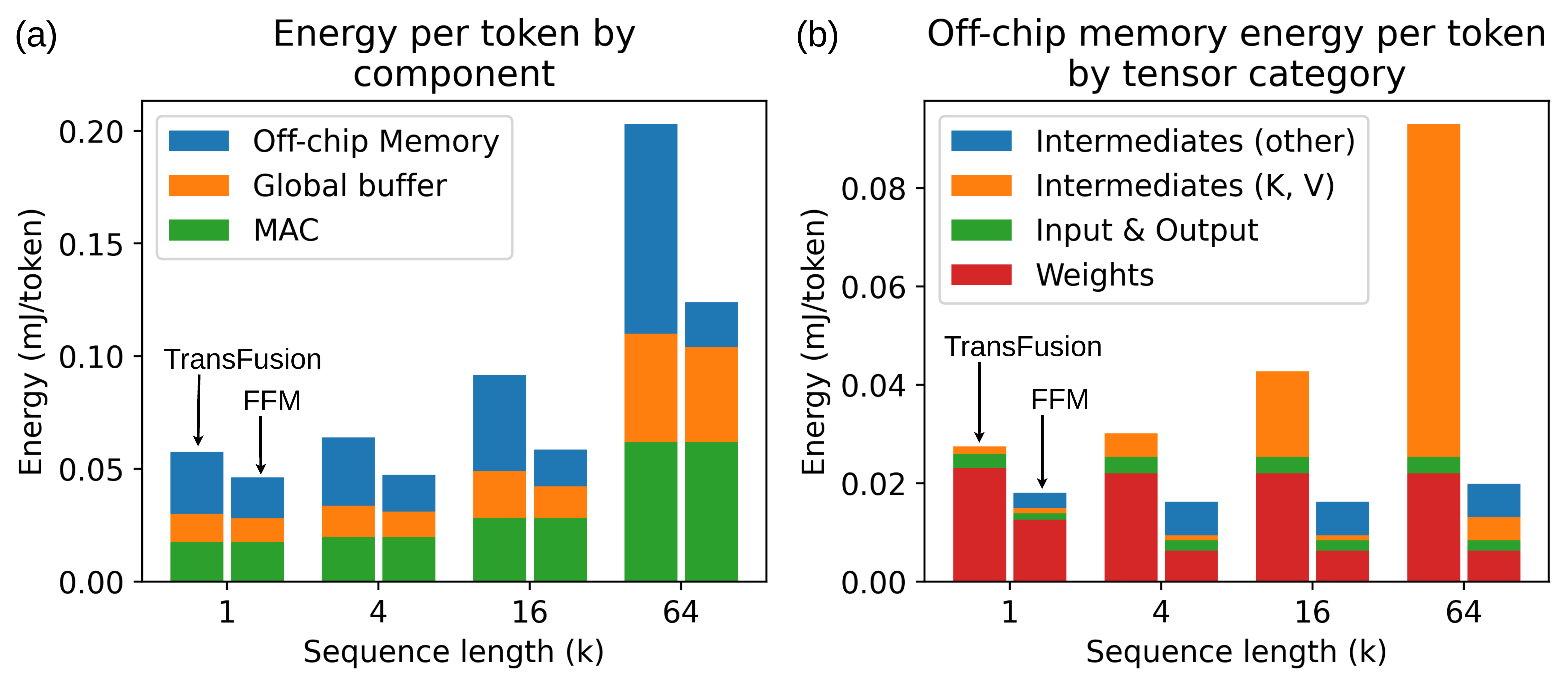}
    \caption{(a) \ffm reduces off-chip memory energy compared to TransFusion. (b) \ffm picks a fusion choice that balances the data movement of various tensors, for a lower overall energy. TransFusion always fuses intermediates other than K and V (notice no ``Intermediates (other)"). }
    \label{fig:energy_breakdown}
\end{figure}

Fig.~\ref{fig:comparison_summary} shows that \textbf{\ffm} improves EDP, energy, and latency relative to \textbf{TransFusion}. Improvements increase at longer sequence lengths (up to 1.8$\times$ lower EDP). From Fig.~\ref{fig:energy_breakdown}(a), which breaks down energy by component, we can see that \ffm reduces off-chip memory energy significantly. Since \textbf{\ffm} benefits from reducing data movement, there is a greater reduction in energy than in latency. This is because reducing data movement always helps energy, but only helps latency until an Einsum becomes compute-bound.

To understand the source of \textbf{\ffm}'s improvements, note that there are two types of reuse opportunities: inter-Einsum (fusion) and intra-Einsum (one Einsum reusing a tensor multiple times). \textbf{\ffm} balances optimizing for both types, while \textbf{TransFusion} primarily focuses on maximizing inter-Einsum reuse by fusing all Einsums except $K$ and $V$, often at the cost of intra-Einsum reuse because they compete for on-chip buffer memory.

The impact of balancing intra- and inter-Einsum reuse can be seen in the off-chip energy breakdown by tensors in Fig.~\ref{fig:energy_breakdown}(b). Note that \textbf{TransFusion}, which focuses on inter-Einsum reuse, always keeps on-chip shared tensors other than $K$ and $V$ (see ``Intermediates (other)"). \textbf{TransFusion} keeps these intermediates on-chip at the cost of intra-Einsum reuse, which suffers from a lack of on-chip buffer capacity, and therefore increasing ``Weights" and ``Intermediates (K, V)" energy. In contrast, \textbf{\ffm} fuses fewer Einsums, which increases data movement of ``Intermediates (other)", but frees on-chip buffer capacity to increase intra-Einsum reuse, keeping ``Weights" and ``Intermediates (K,V)" energy low.

\begin{figure}
    \centering
    \begin{subfigure}{0.9\linewidth}
        \includegraphics[width=\linewidth]{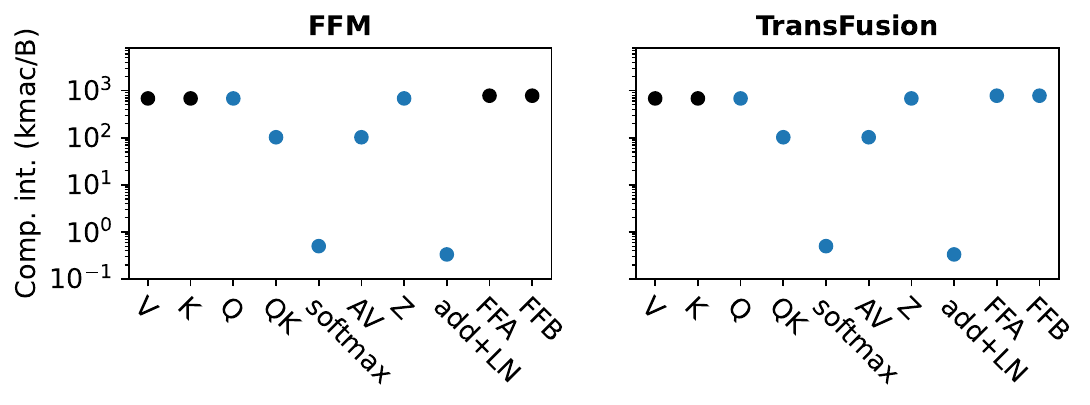}
        \caption{Sequence length = 1k}
    \end{subfigure}
    \begin{subfigure}{0.9\linewidth}
        \includegraphics[width=\linewidth]{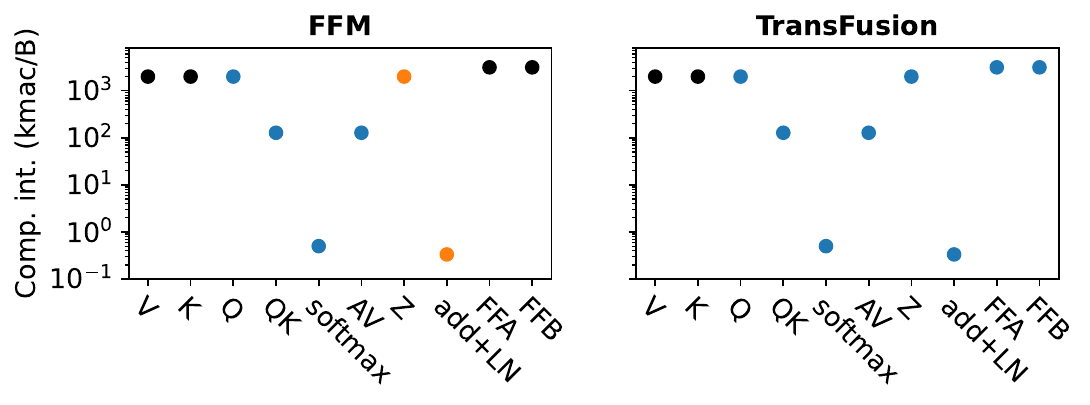}
        \caption{Sequence length = 64k}
    \end{subfigure}
    \caption{FFM strategically fuses Einsums with low compute intensities, which benefits most from fusion. The x-axis shows the Einsums of a GPT-3 layer and the y-axis shows the computational intensity. Unfused Einsums are in black, and different sets of fused Einsums are given different colors. Sequence length is 64k tokens.}
    \label{fig:fusion_set}
\end{figure}

We can gain more insight into how \textbf{\ffm} balances leveraging intra- and inter-Einsum reuse by analyzing which Einsums are fused in the resulting mapping. Fig.~\ref{fig:fusion_set} shows \textbf{\ffm's} fusion choices for GPT-3 with the longest and shortest sequence lengths. We make two observations. First, Fig.~\ref{fig:fusion_set} shows that in both sequence lengths, \textbf{\ffm} prioritizes fusing Einsums with the lowest compute intensity first (\ie low intra-Einsum reuse opportunities), which gains the most benefit from fusion. This choice contrasts with \textbf{TransFusion}, which fuses all Einsums except K and V. As shown before, \textbf{\ffm's} choice, which balances intra- and inter-Einsum reuse, results in lower energy overall.

Second, the optimal balance of intra- and inter-Einsum reuse depends on workload shape, which changes the amount of each reuse type and the costs of achieving them. For example, Fig.~\ref{fig:fusion_set} shows that \textbf{\ffm} does not fuse Einsum $AV$ with $Z$ when the sequence length is longer, keeping the intermediate tensor off-chip instead. This decision is made for two reasons. (1) The size of the intermediate tensor between Einsums $AV$ and $Z$ to keep on-chip increases with sequence length (\ie the cost of fusion is higher). (2) Longer sequence lengths increase the amount of intra-Einsum reuse, lowering the relative benefit of fusion. Given these tradeoffs and limited on-chip capacity, \textbf{\ffm} does not fuse $AV$ and $Z$ at longer sequence length. This is why a single choice, even an optimized one such as \textbf{TransFusion}, will not be optimal in every scenario.

\section{Related Works}
Many prior works propose mappers for single-Einsum workloads~\cite{timeloop, maestro, zigzag, gamma_mapper, mindmappings, cosa}. Multi-Einsum mappers have also been proposed for particular workloads and/or dataflows (commonly only convolutional neural networks)~\cite{defines, convfusion, optimus, fusedcnn}. Most recently, mappers that support a wider range of workloads and a larger mapspace (\eg more dataflow choices) have been proposed~\cite{set, tileflow}. However, these mappers explore their large mapspaces with black-box optimization methods, which we have shown to converge infeasibly slowly.

Optimal mappings generated by FFM can be translated into loop nests, which can guide compiler tool chains~\cite{xla,tvm, taco}.\label{sec:related_works:compilers}

\section{Conclusion}
Driven in part by the rise of deep neural networks and big data, tensor algebra workloads have become critical to accelerate, and even small differences in energy or latency can become great when these workloads are deployed at scale. Moreover, fusion has emerged as a key optimization. However, it is a challenge to find optimal fused mappings due to the exponential mapspace size. The \ffmlong partitions this exploration by introducing the notion of partial mappings and defines attributes (\eg compatibility, resource consumption, costs) that capture how they contribute to full mappings. This lets the \ffmlong quickly find optimal fused mappings in a comprehensive mapspace.



\bibliographystyle{ACM-Reference-Format}
\bibliography{refs}

\end{document}